\documentclass[11pt, twocolumn]{article}
\usepackage{geometry}
\usepackage{graphicx}
\usepackage{amssymb}
\usepackage{amsmath}
\usepackage{epstopdf} 
\usepackage{authblk}
\usepackage{abstract}
\newcommand{\ux}{\underline{x}}
\newcommand{\Sres}{\Sigma_\text{res}}
\newcommand{\Nres}{N_\text{res}}

\title{The Backtracking Survey Propagation Algorithm\\for Solving Random K-SAT Problems}
\author[$1$]{Raffaele Marino}
\affil[$1$]{NORDITA and AlbaNova University Centre, Dept. of Computational Biology, KTH-Royal Institute of Technology and Stockholm University, Roslagstullsbacken 23, SE-10691 Stockholm, Sweden\\
email:~\texttt{rmarino@kth.se}}
\author[$2$]{Giorgio Parisi}
\author[$2$]{Federico Ricci-Tersenghi}
\affil[$2$]{Dipartimento di Fisica, Sapienza Universit\`a di Roma and Istituto Nazionale di Fisica Nucleare, Sezione di Roma1 and CNR-Nanotec, Unit\`a di Roma, P.le Aldo Moro 5, I-00185, Roma, Italy\\
emails:~\texttt{giorgio.parisi@roma1.infn.it}, \texttt{federico.ricci@roma1.infn.it}}
\date{\today}

\begin{document}

\twocolumn[
\maketitle
\begin{onecolabstract}
Discrete combinatorial optimization plays a central role in many scientific disciplines, however for hard problems we lack linear time algorithms that would allow us to solve very large instances. Moreover it is still unclear what are the key features that make a discrete combinatorial optimization problem hard to solve.
Here we study random $K$-satisfiability problems with $K=3,4$ which are known to be very hard close to the SAT-UNSAT threshold, where problems stop having solutions.
We show that the Backtracking Survey Propagation algorithm, in a time practically linear in the problem size, is able to find solutions very close to the threshold, in a region unreachable by any other algorithm.
All solutions found have no frozen variables, thus supporting the conjecture that only unfrozen solutions can be found in linear time, and that a problem becomes impossibile to solve in linear time when all solutions contain frozen variables.
\end{onecolabstract}
]

Optimization problems with discrete variables are widespread among scientific disciplines and often among the hardest to solve.
The $K$-satisfiability ($K$-SAT) problem is a combinatorial discrete optimization problem of $N$ Boolean variables, $\ux=\{x_i\}_{i=1,N}$, submitted to $M$ constraints. Each constraint, called clause, is in the form of an OR logical operator of $K$ literals (variables and their negations): the problem is solvable when there exists at least one configuration of the variables, among the $2^N$ possible ones, that satisfies all constraints. The $K$-SAT problem for $K \geq 3$ is a central problem in combinatorial optimization: it was among the first problems shown to be $NP$-complete \cite{Cook71, bookGareyJohnson, bookPapa} and is still very much studied.
A growing collaboration between theoretical computer scientists and statistical physicists has focused on the random $K$-SAT ensemble \cite{bookMezardMontanari,bookMooreMertens}, where each formula is generated by randomly choosing $M=\alpha N$ clauses of $K$ literals. Formulas from this ensemble become extremely hard to solve when the clause to variable ratio $\alpha$ grows \cite{CookMitch1997}: nevertheless, even in this region, the locally tree-like structure of the factor graph \cite{factorGraph}, representing the interaction network among variables, makes the random $K$-SAT ensemble a perfect candidate for analytic computations.
The study of random $K$-SAT problems and of the related solving algorithms is likely to shed light on the origin of the computational complexity and to allow for the development of improved solving algorithms.

Both numerical \cite{KirkpatrickSelman1994} and analytical \cite{DBM2000,DMSZ2001} evidence suggest that a threshold phenomenon takes place in random $K$-SAT ensembles: in the limit of very large formulas, $N \to \infty$, a typical formula has a solution for $\alpha<\alpha_{\rm s}(K)$, while it is unsatisfiable for $\alpha>\alpha_{\rm s}(K)$.
It has been very recently proved in Ref.~\cite{DingSlySun2014} that for $K$ large enough the SAT-UNSAT threshold $\alpha_{\rm s}(K)$ exists in the $N\to\infty$ limit and coincides with the prediction from the cavity method of statistical physics \cite{MPZ2002}. A widely accepted conjecture is that the SAT-UNSAT threshold $\alpha_{\rm s}(K)$ exists for any value of $K$.
Finding solutions close to $\alpha_{\rm s}$ is very hard, and all known algorithms running in polynomial time fail to find solutions when $\alpha>\alpha_{\rm a}$, for some $\alpha_{\rm a} < \alpha_{\rm s}$. Actually, each algorithm ALG has it own algorithmic threshold $\alpha_{\rm a}^\text{\tiny ALG}$, such that the probability of finding a solution vanishes for $\alpha > \alpha_{\rm a}^\text{\tiny ALG}$ in the large $N$ limit.
For most algorithms $\alpha_{\rm a}^\text{\tiny ALG}$ is well below $\alpha_{\rm s}$.
We define $\alpha_{\rm a} = \max_\text{\tiny ALG} \alpha_{\rm a}^\text{\tiny ALG}$ the threshold beyond which no polynomial-time algorithm can find solutions.
There are two main open questions: to find improved algorithms having a larger $\alpha_{\rm a}^\text{\tiny ALG}$, and to understand what is the theoretical upper bound $\alpha_{\rm a}$.
Here we present progress on both issues.

The best prediction about the SAT-UNSAT threshold comes from the cavity method \cite{MPZ2002,MezardZecchina2003,MMZ2006,MRTS2008}: for example, $\alpha_{\rm s}(K\!=\!3)\!=\!4.2667$ \cite{MMZ2006} and $\alpha_{\rm s}(K\!=\!4)\!=\!9.931$ \cite{MRTS2008}.
Actually the statistical physics study of random $K$-SAT ensembles also provides us with a very detailed description of how the space of solutions changes when $\alpha$ spans the whole SAT phase ($0\le\alpha\le\alpha_{\rm s}$).
Let us consider typical formulas in the large $N$ limit and the vast majority of solutions in these formulas (i.e.\ typical solutions), we know that, at low enough $\alpha$ values, the set of solutions is connected, so that they form a single cluster. In SAT problems we say 2 solutions are neighbors if they differ in the assignment of just one variable; in other problems (e.g.\ XOR-SAT \cite{MRTZ2003}) this definition of neighbor needs to be relaxed, because a pair of solutions differing in just one variable are not allowed by the model definition. As long as the notion of neighborhood is relaxed to Hamming distances $o(N)$ all the picture of the solution space based on statistical physics remains unaltered.

As $\alpha$ increases, not only the number of solutions decreases, but at $\alpha_{\rm d}$ the random $K$-SAT ensemble  undergoes a phase transition: the space of solutions shatters into an exponentially large (in the problem size $N$) number of clusters; two solutions belonging to different clusters have a Hamming distance $O(N)$. 
If we define the energy function $E(\ux)$ as the number of unsatisfied clauses in configuration $\ux$, it has been found \cite{MPZ2002} that for $\alpha>\alpha_{\rm d}$ the energy $E(\ux)$ has exponentially many (in $N$) local minima of positive energy, which may trap algorithms that look for solutions by energy relaxation (e.g. Monte Carlo simulated annealing).

Further increasing $\alpha$, each cluster loses solutions and shrinks, but the most relevant change is in the {\bf number} of clusters. The cavity method allows us to count clusters of solutions as a function of the number of solutions they contain \cite{KMRTSZ2007}: using this very detailed description several other phase transitions have been identified \cite{ZK2007,MRTS2008}.
For example, there is a value $\alpha_{\rm c}$ where a condensation phase transition takes place, such that for $\alpha>\alpha_{\rm c}$ the vast majority of solutions belong to a sub-exponential number of clusters, leading to effective long-range correlations among variables in typical solutions, which are hard to approximate by any algorithm with a finite horizon. In general $\alpha_{\rm d} \le \alpha_{\rm c} \le \alpha_{\rm s}$ holds.
Most of the above picture of the solution space has been proven rigorously in the large $K$ limit \cite{ACO2008,ACORT2011}.

Moving to the algorithmic side, a very interesting question is whether such a rich structure of the solution space affects the performance of searching algorithms. While clustering at $\alpha_{\rm d}$ may have some impact on algorithms that sample solutions uniformly \cite{RTS2009}, many algorithms exist that can find at least one solution with $\alpha>\alpha_{\rm d}$ \cite{MPZ2002,FMS,ASAT}.

A solid conjecture is that the hardness of a formula is related to the existence of a subset of highly correlated variables, which are very hard to assign correctly altogether; the worst case being a subset of variables that can have a {\bf unique} assignment. This concept was introduced with the name of backbone in Ref.~\cite{RemiRiccardoNature}. The same concept applied to solutions within a single cluster lead to the  definition of {\bf frozen variables} (within a cluster) as those variables taking the same value in all solutions of the cluster \cite{SemerjianFreezing}.
It has been proven in Ref.~\cite{maneva2007new} that the fraction of frozen variables in a cluster is either zero or lower bounded by $(\alpha e^2)^{-1/(K-2)}$; in the latter case the cluster is called frozen.

According to the above conjecture, finding a solution in a frozen cluster is hard (in practice it should require a time growing exponentially with $N$). So the smartest algorithm running in polynomial time should search for unfrozen clusters as long as they exist.
Unfortunately counting unfrozen clusters is not an easy job, and indeed a large deviation analysis of their number has been achieved only very recently \cite{SigmaUnfrozen} for a different and simpler problem (bicoloring random regular hypergraphs).
For random $K$-SAT only partial results are known, that can be stated in terms of two thresholds: for $\alpha>\alpha_{\rm r}$ (rigidity) {\bf typical} solutions are in frozen cluster (but a minority of solutions may still be unfrozen), while for $\alpha>\alpha_{\rm f}$ (freezing) {\bf all} solutions are frozen.
It has been rigorously proven \cite{AchRT2006,AchRT2009} that $\alpha_{\rm f}<\alpha_{\rm s}$ holds strictly for $K>8$.
For small $K$, which is the interesting case for benchmarking solving algorithms, we know $\alpha_{\rm r} = 9.883(15)$ for $K=4$ from the cavity method \cite{MRTS2008}, while for $K=3$ the estimate $\alpha_{\rm f}=4.254(9)$ comes from exhaustive enumerations in small formulas ($N\le 100$) \cite{ArdeliusZdeborova} and is likely to be affected by strong finite size effects.
In general $\alpha_{\rm d} \le \alpha_{\rm r} \le \alpha_{\rm f} \le \alpha_{\rm s}$ holds.

The conjecture above implies that no polynomial time algorithm can solve problems with $\alpha\ge\alpha_{\rm f}$, but also finding solutions close to the rigidity threshold $\alpha_{\rm r}$ is expected to be very hard, given that unfrozen solutions becomes a tiny minority. And this is indeed what happens for all known algorithms.
Since we are interested in solving very large problems we only consider algorithms whose running time scales almost linearly with $N$ and we measure performance of each algorithm in terms of its algorithmic threshold $\alpha_{\rm a}^\text{\tiny ALG}$.

Solving algorithms for random $K$-SAT problems can be roughly classified in two main categories: algorithms that {\bf search} for a solution by performing a biased random walk in the space of configurations and algorithms that try to {\bf build} the solutions by assigning variables, according to some estimated marginals.
WalkSat \cite{WalkSat}, focused Metropolis search (FMS) \cite{FMS} and ASAT \cite{ASAT} belong to the former category; while in the latter category we find Belief Propagation guided Decimation (BPD) \cite{RTS2009} and Survey Inspired Decimation (SID) \cite{BMZ2005}.
All these algorithms are rather effective in finding solutions to random $K$-SAT problems: e.g.\ for $K\!=\!4$ we have $\alpha_{\rm a}^\text{\tiny BPD}=9.05$, $\alpha_{\rm a}^\text{\tiny FMS}\simeq 9.55$ and $\alpha_{\rm a}^\text{\tiny SID}\simeq 9.73$ to be compared with a much lower algorithmic threshold $\alpha_{\rm a}^\text{\tiny GUC}\!= 5.54$ achieved by Generalized Unit Clause, the best algorithm whose range of convergence to a solution can be proven rigorously \cite{FriezeSuen1996}.
Among the efficient algorithms above, only BPD can be solved analytically \cite{RTS2009} to find the algorithmic threshold $\alpha_{\rm a}^\text{\tiny BPD}$; for the others we are forced to run extensive numerical simulations to measure $\alpha_{\rm a}^\text{\tiny ALG}$.

At present the algorithm achieving the best performance on several constraint satisfaction problems is SID, which has been successfully applied to the random $K$-SAT problem \cite{MPZ2002} and to the coloring problem \cite{MPWZ2002}.
The statistical properties of the SID algorithm for $K\!=\!3$ have been studied in details in Refs.~\cite{ParisiRemarks2003,BMZ2005}. 
Numerical experiments on random $3$-SAT problems with a large number of variables, up to $N=3\times 10^5$, show that in a time that is approximately linear in $N$ the SID algorithm finds solutions up to $\alpha_{\rm a}^\text{\tiny SID} \simeq 4.2525$ \cite{ParisiRemarks2003}, that is definitely smaller, although very close to, $\alpha_{\rm s}(K\!=\!3)=4.2667$. In the region $\alpha_{\rm a}^\text{\tiny SID}< \alpha< \alpha_{\rm s} $ the problem is satisfiable for large $N$, but at present no algorithm can find solutions there.

To fill this gap we study a new algorithm for finding solutions to random $K$-SAT problems, the Backtracking Survey Propagation (BSP) algorithm. This algorithm (fully explained in the Methods section) is based, as SID, on the survey propagation (SP) equations derived within the cavity method \cite{MPZ2002,ParisiRemarks2003,BMZ2005} that provide an estimate on the total number of clusters $\mathcal{N}_\text{clus}=\exp(\Sigma)$.
The BSP algorithm, like SID, aims at assigning gradually the variables such as to keep the complexity $\Sigma$ as large as possibile, i.e.\ trying not to kill too many clusters \cite{ParisiRemarks2003}.
While in SID each variable is assigned only once, in BSP we allow unsetting variables already assigned such as to backtrack on previous non-optimal choices.
In BSP the $r$ parameter is the ratio between the number of backtracking moves (unsetting one variable) and the number of decimation moves (assigning one variable). $r< 1$ must hold and for $r=0$ we recover the SID algorithm. The running time scales as $N/(1-r)$, with a slight overhead for maintaining the data structures, making the running time effectively linear in $N$ for any $r < 1$.

The idea supporting backtracking \cite{ParisiBSP2003} is that a choice made at the beginning of the decimation process, when most of the variables are unassigned, may turn to be suboptimal later on; if we re-assign a variable that is no longer consistent with the current best estimate of its marginal probability, we may get a better satisfying configuration.
We do not expect the backtracking to be essential when correlations between variables are short ranged, but approaching $\alpha_{\rm s}$ we know that correlations become long ranged and thus the assignment of a single variable may affect a huge number of other variables: this is the situation when we expect the backtracking to be crucial.

This idea may look similar in spirit to the survey propagation reinforcement (SPR) algorithm \cite{Chavas2005}, where variables are allowed to change their most likely value during the run, but in practice BSP works much  better. In SPR, once reinforcement fields are large, the re-assignment of any variable becomes unfeasible, while in BSP variables can be re-assigned to better values until the very end, and this is a major advantage.

\section*{Results}

\begin{figure}[t]
\centering
\includegraphics[width=\columnwidth]{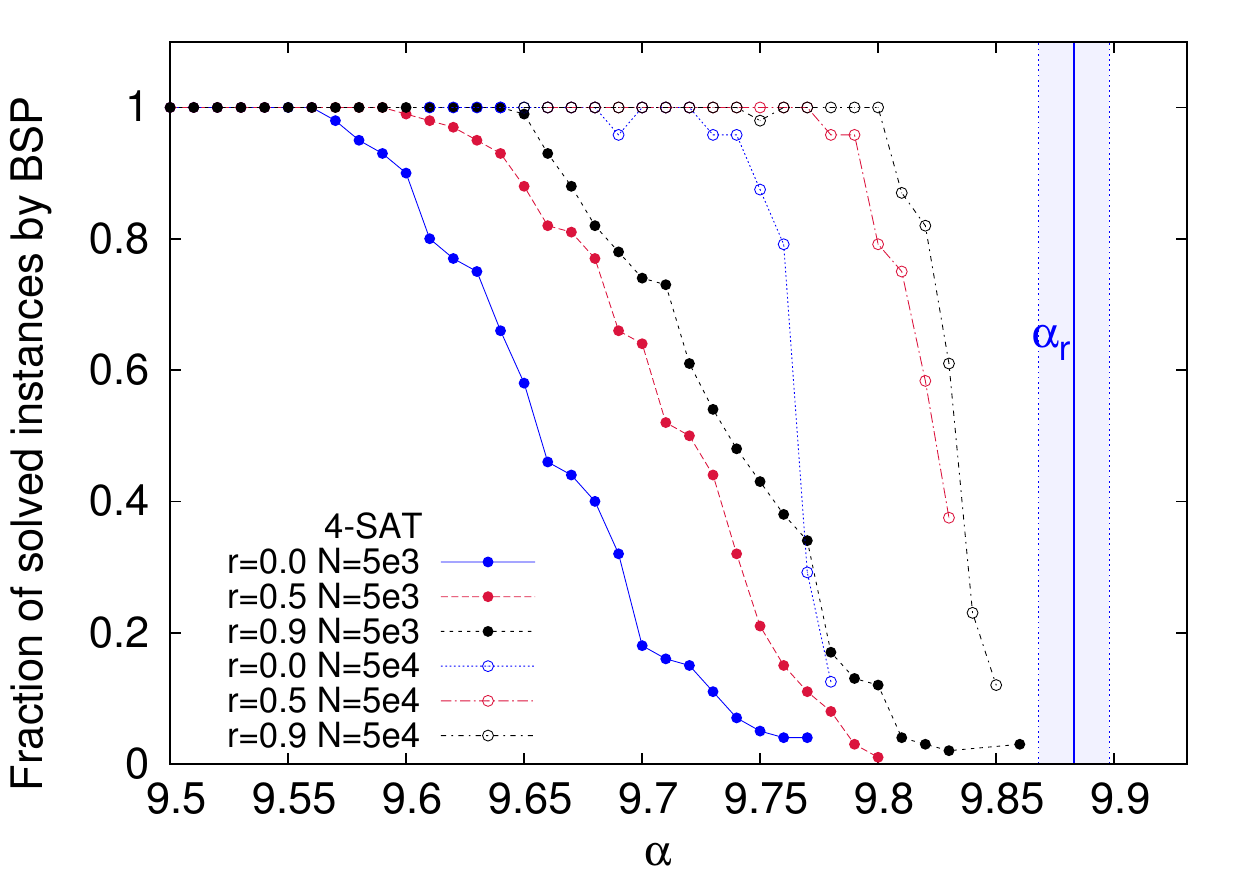}
\caption{\textbf{Fraction of random $\mathbf{4}$-SAT instances solved by BSP} as a function of the constraints per variable ratio $\alpha$. The average is computed over 100 instances with $N=5\,000$ (solid symbols) and $N=50\,000$ (empty symbols) variables.
The vertical line is the best estimate for $\alpha_{\rm r}$ and the shaded region is the statistical error on this estimate. For each instance, the algorithm has been run once; on instances not solved on the first run, a second run rarely ($<1\%$) finds a solution.
The plot shows that the backtracking ($r>0$) definitely makes the BSP algorithm more efficient in finding solutions.
Although data become sharper by increasing the problem size $N$, a good estimation of the algorithmic threshold from these datasets is unfeasible.}
\label{fig1}
\end{figure}

\textbf{Probability of finding a SAT assignment} The standard way to study the performance of a solving algorithm is to measure the fraction of instances it can solve as a function of $\alpha$. We show in Fig.~\ref{fig1} such a fraction for BSP run with three values of the $r$ parameter ($r=0,\,0.5$ and $0.9$) on random $4$-SAT problems of two different sizes ($N=5\,000$ and $N=50\,000$).
The probability of finding a solution increases both with $r$ and $N$, but an extrapolation to the large $N$ limit of these data is unlikely to provide a reliable estimation of the algorithmic threshold $\alpha_{\rm a}^\text{\tiny BSP}$.

In each plot having $\alpha$ on the abscissa, the right end of the plot coincides with the best estimate of $\alpha_{\rm s}$, in order to provide an immediate indication of how close to the SAT-UNSAT threshold the algorithm can work.

\begin{figure}[!t]
\centering
\includegraphics[width=\columnwidth]{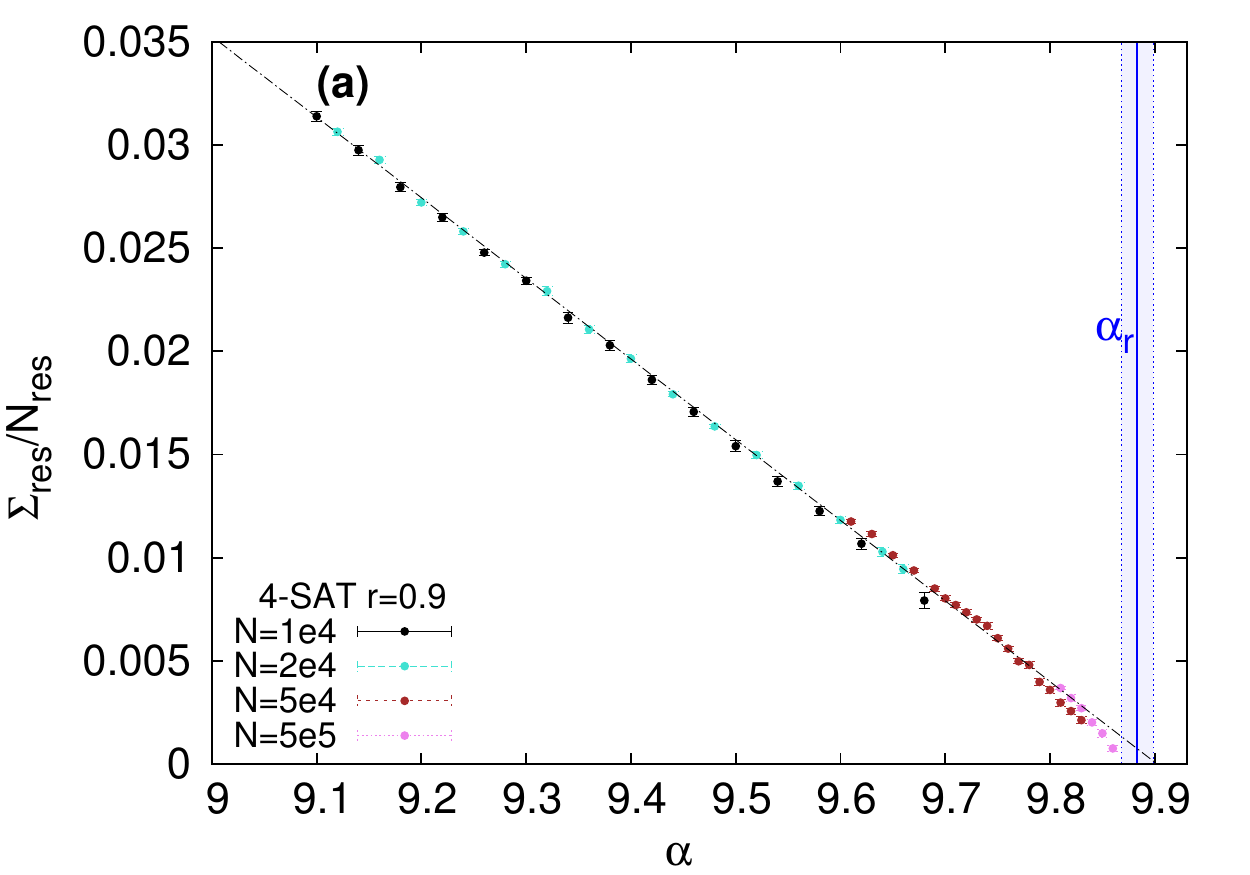}
\includegraphics[width=\columnwidth]{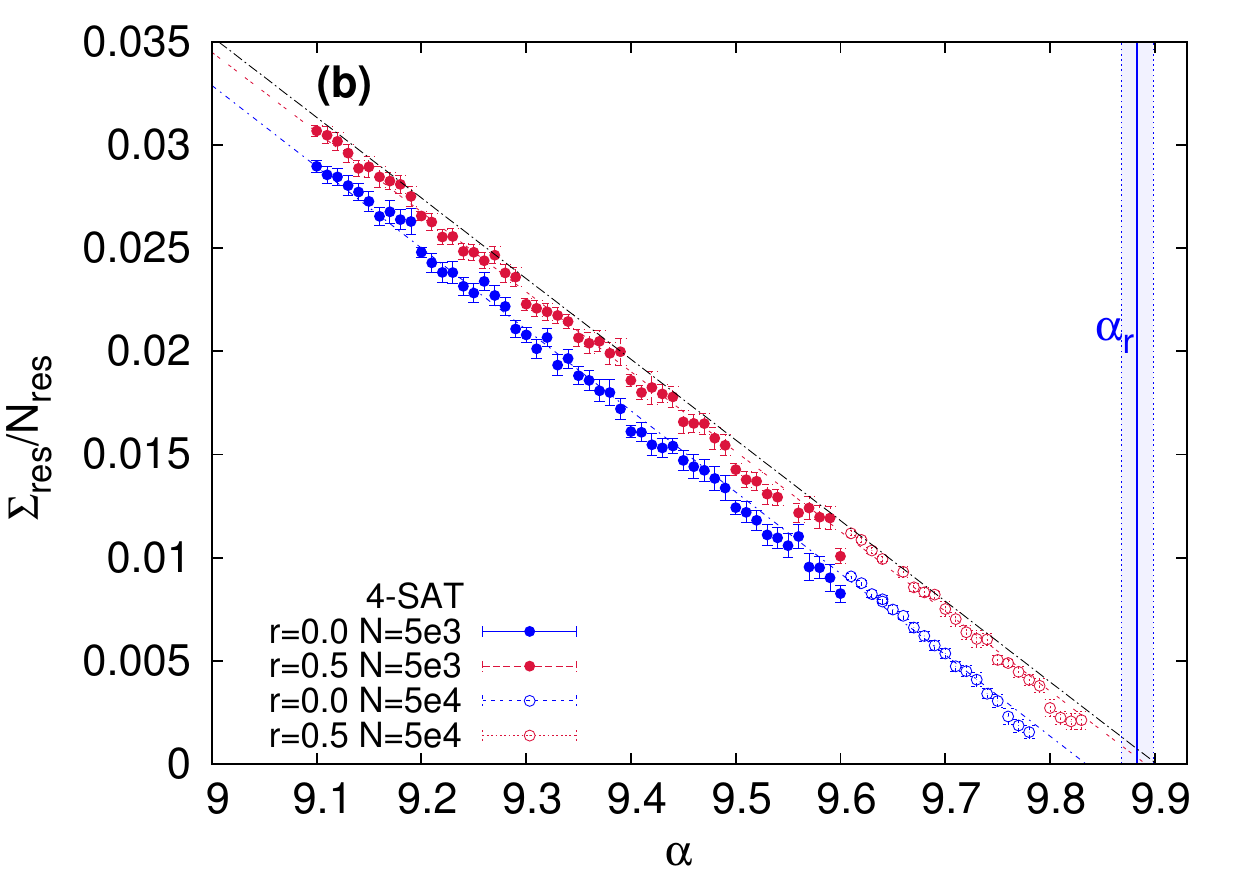}
\caption{\textbf{BSP algorithmic threshold on random $4$-SAT problems}. The residual complexity per variable, $\Sres/\Nres$, goes to zero at the algorithmic threshold $\alpha_{\rm a}^\text{\tiny BSP}$. (\textbf{a}) The very small finite size effects, mostly producing a slight downward curvature at the right end, allow for a very reliable estimate of $\alpha_{\rm a}^\text{\tiny BSP}$ via a linear fit. For random $4$-SAT problems solved by BSP with $r=0.9$ we get $\alpha_{\rm a}^\text{\tiny BSP}\approx 9.9$, slightly beyond the rigidity threshold $\alpha_{\rm r}=9.883(15)$, marked by a vertical line (the shaded area being its statistical error). (\textbf{b}) The same linear extrapolation holds for other values of $r$ (red dotted line for $r=0.5$ and blue dashed line for $r=0$). The black line is the fit to $r=0.9$ data shown in panel (\textbf{a}). SID without backtracking ($r=0$) has a much lower algorithmic threshold, $\alpha_{\rm a}^\text{\tiny SID}\approx 9.83$. Error bars in both panels are the standard error on the mean (sem).}
\label{fig2}
\end{figure}

\vspace{5mm}\noindent\textbf{Order parameter and algorithmic threshold} In order to obtain a reliable estimate of $\alpha_{\rm a}^\text{\tiny BSP}$ we look for an order parameter vanishing at $\alpha_{\rm a}^\text{\tiny BSP}$ and having very little finite size effects.
We identify this order parameter with the quantity $\Sres/\Nres$, where $\Sres$ and $\Nres$ are respectively the complexity (i.e.\ log of number of clusters) and the number of unassigned variables in the {\bf residual formula}. As explained in Methods, BSP assigns and re-assigns variables, thus modifying the formula, until the formula simplifies enough that the SP fixed point has only null messages: the residual formula is defined as the last formula with non-null SP fixed point messages.
We have experimentally observed that the BSP algorithm (as the SID one \cite{ParisiRemarks2003}) can simplify the formula enough to reach the trivial SP fixed point only if the complexity $\Sigma$ remains strictly positive during the whole decimation process. In other words, on every run where $\Sigma$ becomes very close to zero or negative, SP stops converging or a contradiction is found. This may happen either because the original problem was unsatisfiable or because the algorithm made some wrong assignments incompatible with the few available solutions.
Thanks to the above observation we have that $\Sres\ge 0$ and thus a null value for the mean residual complexity signals that the BSP algorithm is not able to find any solution, and thus provides a valid estimate for the algorithmic threshold $\alpha_{\rm a}^\text{\tiny BSP}$.
From the statistical physics solution to random $K$-SAT problems we expect $\Sres$ to vanish linearly in $\alpha$.

As we see in panel (a) of Fig.~\ref{fig2} the mean value of the intensive mean residual complexity $\Sres/\Nres$ is practically size-independent and a linear fit provides a very good data interpolation: tiny finite size effects are visible in the largest $N$ datasets only close to the dataset right end.
The linear extrapolation predicts $\alpha_{\rm a}^\text{\tiny BSP}\approx 9.9$ (for $K\!=\!4$ and $r\!=\!0.9$), which is slightly above the rigidity threshold $\alpha_{\rm r}=9.883(15)$ computed in Ref.~\cite{MRTS2008} and reported in the plot with a shaded region corresponding to its statistical error (the value of $\alpha_{\rm f}$ in this case is not known, but $\alpha_{\rm a}^\text{\tiny BSP} < \alpha_{\rm f} \le \alpha_{\rm s}$ should hold).
Although for the finite sizes studied no solution has been found beyond $\alpha_{\rm r}$, Fig.~\ref{fig2} suggests that in the large $N$ limit BSP may be able to find solutions in a region of $\alpha$ where the majority of solutions is in frozen clusters and thus very hard to find. We show below that BSP actually finds solutions in atypical unfrozen clusters, as it has been observed for some smart algorithms solving other kind of constraint satisfaction problems \cite{DallAstaRamezZecchina,LockedCSP}.

The effectiveness of the backtracking can be appreciated in panel (b) of Fig.~\ref{fig2}, where the order parameter $\Sres/\Nres$ is shown for $r=0$ and $r=0.5$, together with linear fits to these datasets and to the $r=0.9$ dataset (black line).
We observe that the algorithmic threshold for BSP is much larger (on the scale measuring the relative distance from the SAT-UNSAT threshold) that the one for SID (i.e.\ $r=0$ dataset).

\begin{figure}[!t]
\centering
\includegraphics[width=\columnwidth]{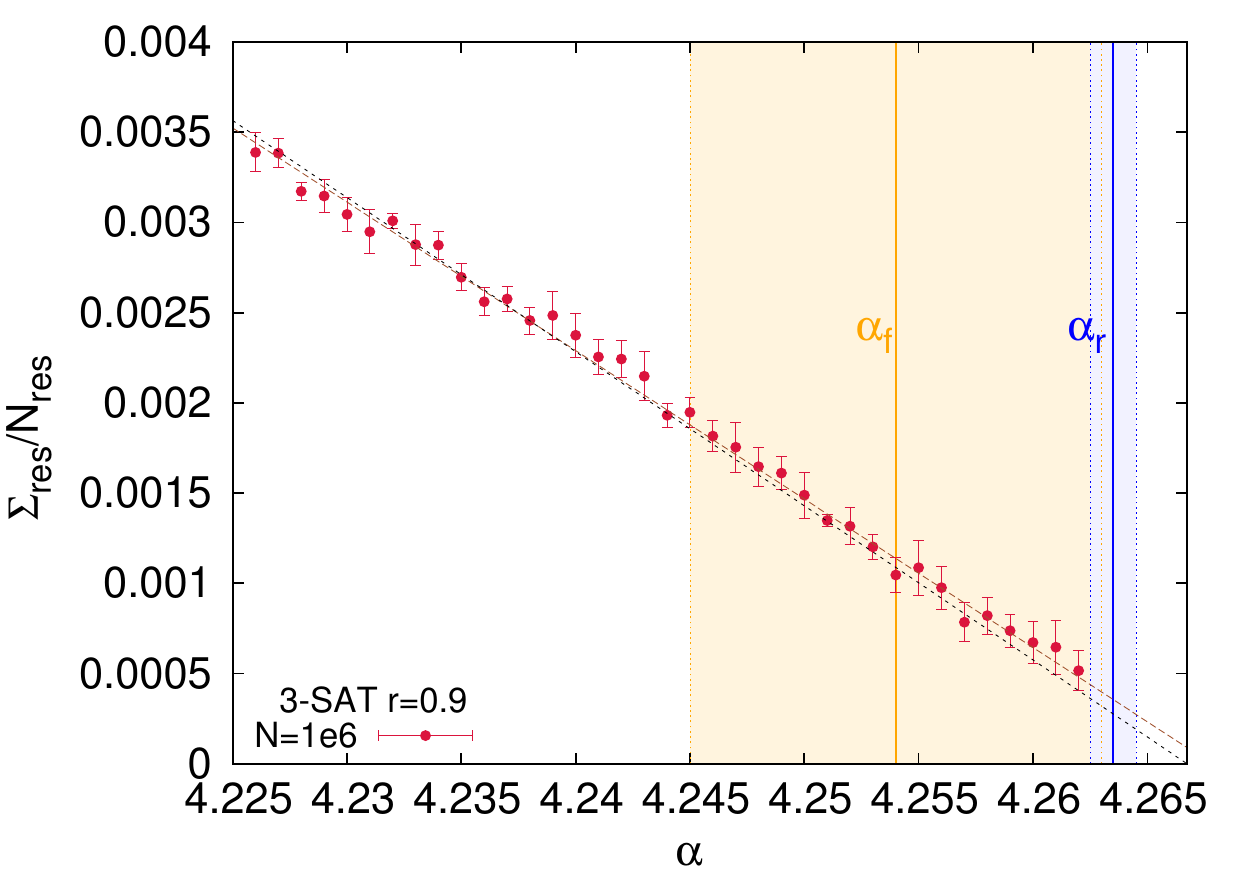}
\caption{\textbf{BSP algorithmic threshold on random $3$-SAT problems}. Same as Fig.~\ref{fig2} for  $K\!=\!3$. The estimate for the freezing threshold $\alpha_{\rm f}=4.254(9)$ measured on small problems in Ref.~\cite{ArdeliusZdeborova} is not very meaningful, given our new estimate for the rigidity threshold $\alpha_{\rm r} = 4.2635(10)$, and the observation that all solutions found by BSP are not frozen. Shaded areas are the statistical uncertainties on the thresholds. A linear fit to the residual complexity (brown line) extrapolates to zero slightly beyond the SAT-UNSAT threshold, at $\alpha_{\rm a}^\text{\tiny BSP}\approx 4.268$, strongly suggesting BSP can find solutions in the entire SAT phase for $K\!=\!3$ in the large $N$ limit. The black line is a linear fit vanishing at $\alpha_{\rm s}$. Error bars are sem.
}
\label{fig3}
\end{figure}

For random $3$-SAT the algorithmic threshold of BSP, run with $r=0.9$, practically coincide with the SAT-UNSAT threshold $\alpha_{\rm s}$ (see Fig.~\ref{fig3}), thus providing a strong evidence that BSP can find solutions in the entire SAT phase.
The estimate for the freezing threshold $\alpha_{\rm f}=4.254(9)$ obtained in Ref.~\cite{ArdeliusZdeborova} from $N\le 100$ data is likely to be too small and affected by strong finite size effects, given that all solutions found by BSP for $N=10^6$ are unfrozen, even beyond the estimated $\alpha_{\rm f}$.
Moreover we have estimated $\alpha_{\rm r} = 4.2635(10)$ improving the data of Ref.~\cite{MRTS2008} and the inequality $\alpha_{\rm r} \le \alpha_{\rm f} \le \alpha_{\rm s}$ makes the above estimate for $\alpha_{\rm f}$ not very meaningful.

\begin{figure}[!ht]
\centering
\includegraphics[width=\columnwidth]{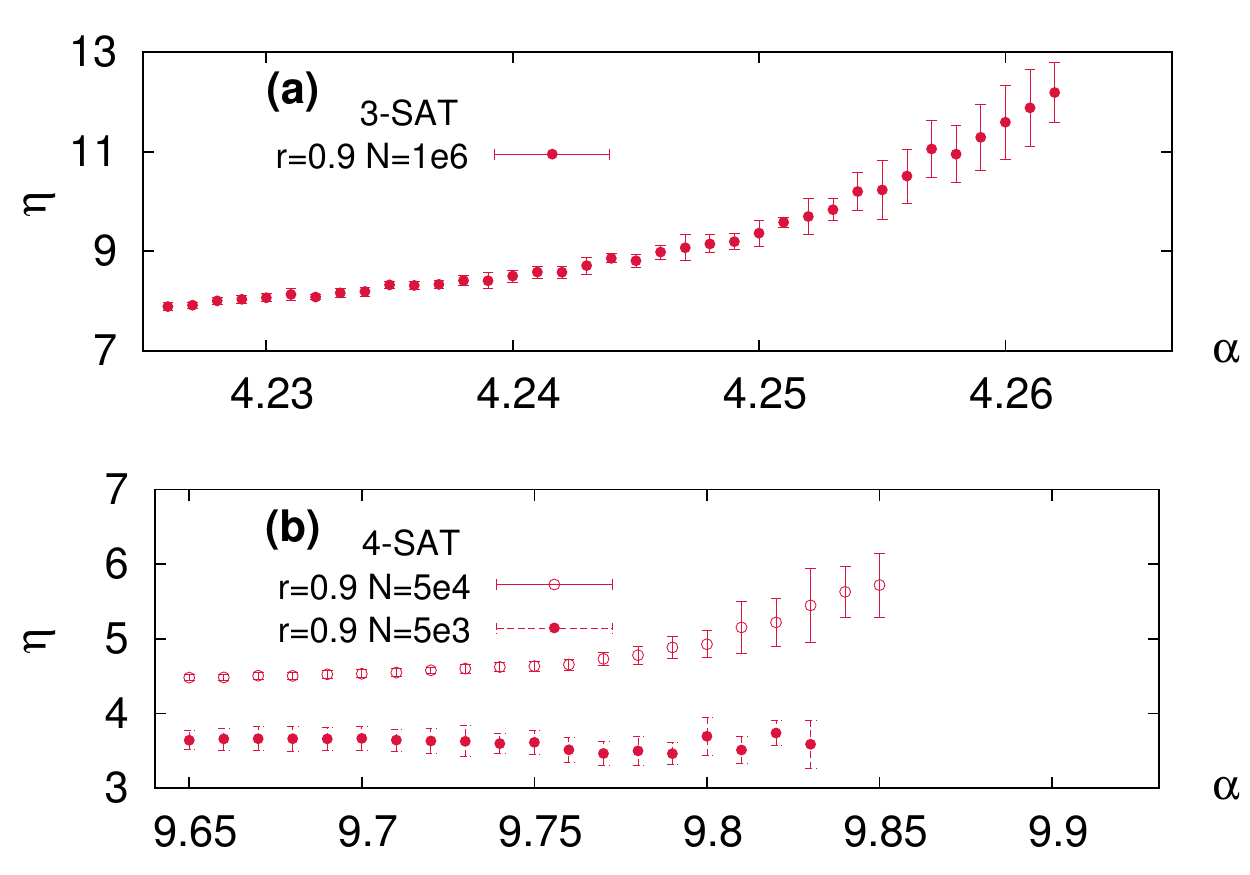}
\caption{\textbf{BSP convergence time}. The mean number $\eta$ of iterations to reach a fixed point of SP equations grows very mildly with $\alpha$ and $N$, both for $K\!=\!3$ (\textbf{a}) and $K\!=\!4$ (\textbf{b}). Error bars are standard deviations.}
\label{fig4}
\end{figure}

\textbf{Computational complexity} As explained in Methods, the BSP algorithm performs $f^{-1}(1-r)^{-1}$ steps roughly, where at each step $fN$ variables are either assigned [with prob.\ $1/(1+r)$] or released [with prob.\ $r/(1+r)$]. At the beginning of each step, the algorithm solves the SP equations with a mean number $\eta$ of iterations. The average $\eta$ is computed only on instances where SP always converges, as is usually done for incomplete algorithms (on the remaining problems the number of iterations reaches the upper limit set by the user, and then BSP exit, returning failure).
Fig.~\ref{fig4} shows that $\eta$ is actually a small number changing mildly with $\alpha$ and $N$ both for $K\!=\!3$ and $K\!=\!4$. The main change that we observe is in the fluctuations of $\eta$ that become much larger approaching $\alpha_{\rm s}$. We expect $\eta$ to eventually grow as $O(\log(N))$, but for the sizes studied we do not observe such a growth.

After convergence to a fixed point, the BSP algorithm just need to sort local marginals, thus the total number of elementary operations to solve an instance grows as $f^{-1}(1-r)^{-1}(a_1\eta N + a_2 N \log N)$, where $a_1$ and $a_2$ are constants.
Moreover, given that the sorting of local marginals does not need to be strict (i.e.\ a partial sorting \cite{partialSorting} running in $O(N)$ time can be enough), we have that in practice the algorithm runs in a time almost linear in the problem size $N$.

\textbf{Whitening procedure} Given that the BSP algorithm is able to find solutions even very close to the rigidity threshold $\alpha_{\rm r}$, it is natural to check whether these solutions have frozen variables or not.
We concentrate on solutions found for random $3$-SAT problems with $N=10^6$, since the large size of these problems makes the analysis very clean.

On each solution found we run the \textit{whitening procedure} (first introduced in \cite{ParisiWhitening} and deeply discussed in \cite{AlfredoRiccardo2004,maneva2007new}), that identifies frozen variables by assigning the joker state $\star$ to unfrozen (white) variables, i.e.\ variables that can take more than one value without violating any clause and thus keeping the formula satisfied. At each step of the whitening procedure, a variables is considered unfrozen (and thus assigned to $\star$) if it belongs only to clauses which either involve a $\star$ variable or are satisfied by another variable. The procedure is continued until all variables are $\star$ or a fixed point is reached: non-$\star$ variables at the fixed point correspond to frozen variables in the starting solution.

We uncover that {\bf all} solutions found by BSP are converted to all-$\star$ by running the whitening procedure, thus showing that solutions found by BSP have no frozen variables. This is somehow expected, according to the conjecture discussed in the Introduction: finding solutions in a frozen cluster would take an exponential time, and so the BSP algorithm actually finds solutions at very large $\alpha$ values by smartly focusing on the sub-dominant unfrozen clusters.

\begin{figure}
\centering
\includegraphics[width=0.85\columnwidth]{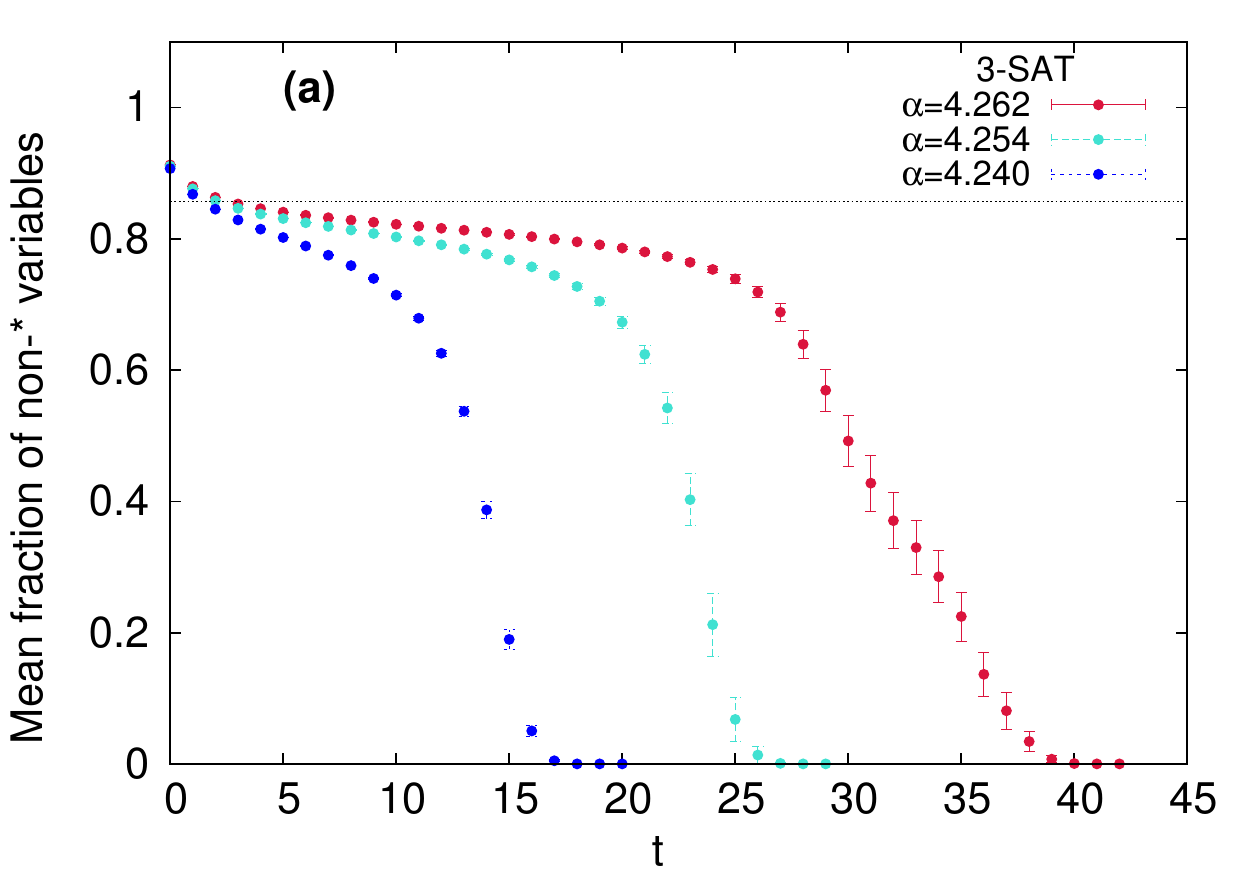}
\includegraphics[width=0.85\columnwidth]{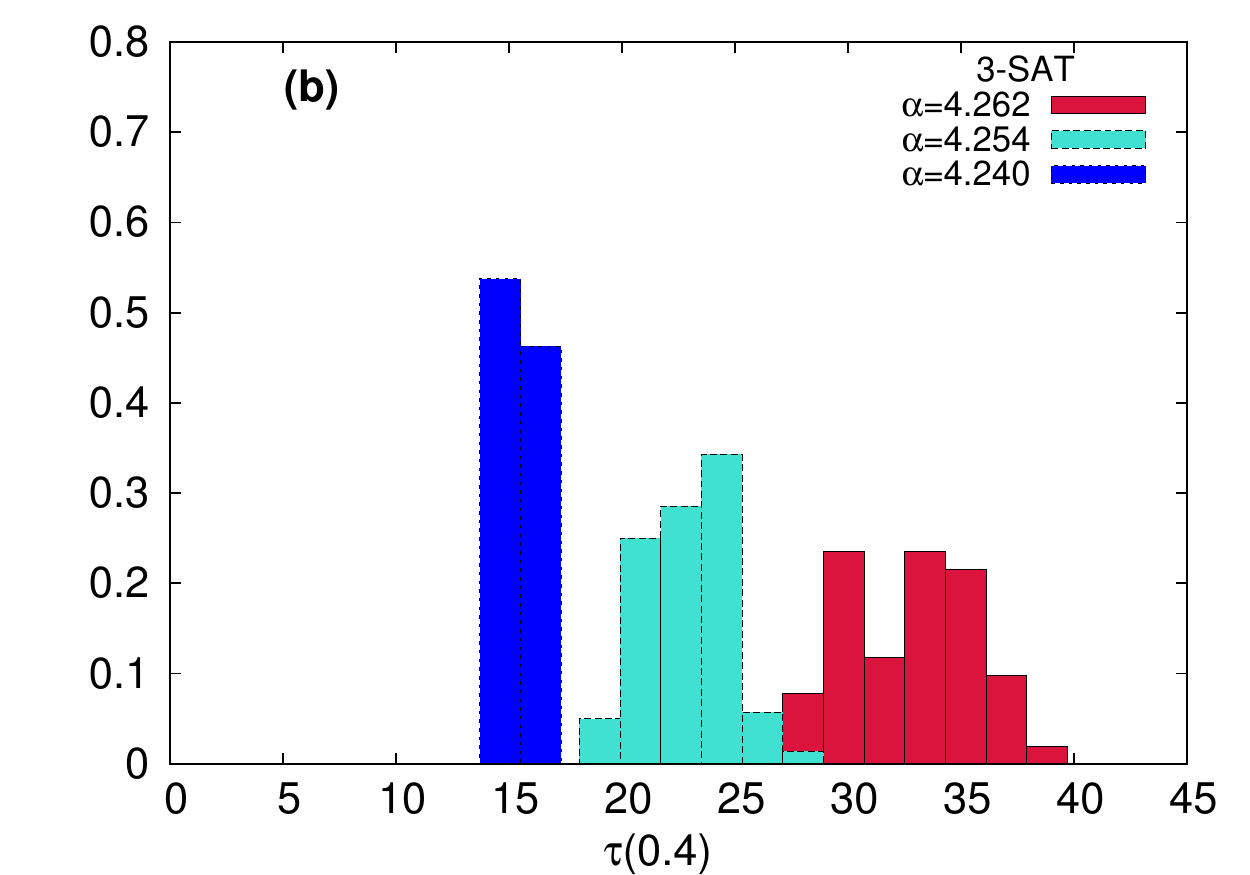}
\includegraphics[width=0.85\columnwidth]{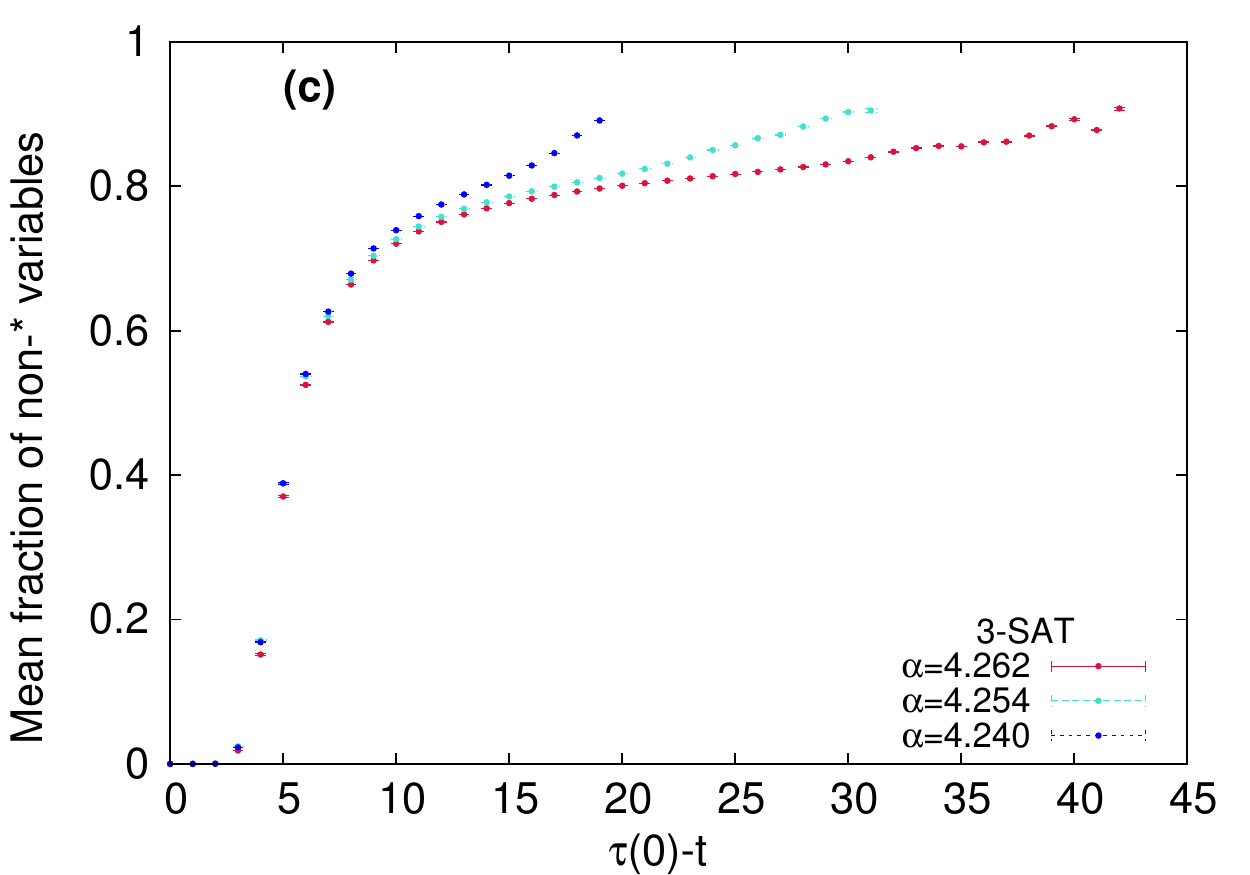}
\caption{\textbf{Whitening random $3$-SAT solutions}. (\textbf{a}) The mean fraction of non-$\star$ variables during the whitening procedure applied to all solutions found by the BSP algorithm goes to zero, following a two steps process. The relaxation time grows increasing $\alpha$ towards the algorithmic threshold. The horizontal line is the SP prediction for the fraction of frozen variables in typical solutions at $\alpha=4.262$ and the comparison with the data shows that solutions found by BSP are atypical. (\textbf{b}) Histograms of the whitening times, defined as the number of iterations required to reach a fraction $0.4$ of non-$\star$ variables. Increasing $\alpha$ both the mean value and the variance of the whitening times grow. (\textbf{c}) Averaging the fraction of non-$\star$ variables at fixed $\tau(0)\!-\!t$, i.e.\ fixing the time to the all-$\star$ fixed point, we get much smaller errors than in panel (\textbf{a}), suggesting that the whitening procedure is practically solution-independent once the plateau is left. In all the panels, error bars are sem.}
\label{fig5}
\end{figure}

The whitening procedure leads to a relaxation of the number of non-$\star$ variables as a function of the number of iterations $t$ that follows a two steps relaxation process \cite{SemerjianFreezing} with an evident plateau, see panel (a) in Fig.~\ref{fig5}, that becomes longer increasing $\alpha$ towards the algorithmic threshold. The time for leaving the plateau, scales as the time $\tau(c)$ for reaching a fraction $c$ on non-$\star$ variables (with $c$ smaller than the plateau value). The latter has large fluctuations from solution to solution, as shown in panel (b) of Fig.~\ref{fig5} for $c=0.4$ (very similar, but shifted, histograms are obtained for other $c$ values). However, after leaving the plateau, the dynamics of the whitening procedure is the same for each solution. Indeed plotting the mean fraction of non-$\star$ variables as a function of the time to reach the all-$\star$ configuration, $\tau(0)-t$, we see that fluctuations are strongly suppressed and the relaxation is the same for each solution (see panel (c) in Fig.~\ref{fig5}).

\begin{figure}[!t]
%\centering
\includegraphics[width=0.95\columnwidth]{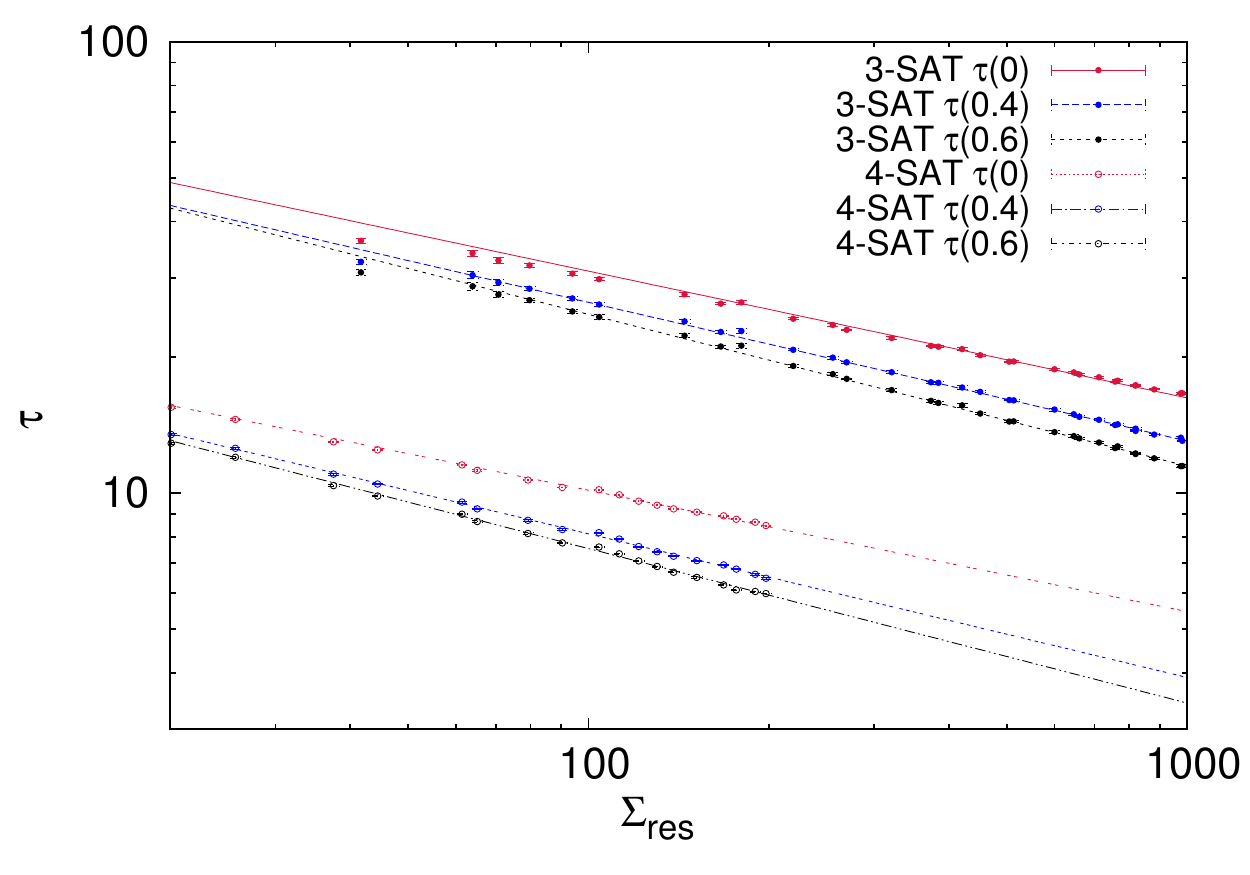}
\caption{\textbf{Critical exponent for the whitening time divergence}. The whitening time $\tau(c)$, defined as the mean time needed to reach a fraction $c$ of non-$\star$ variables in the whitening procedure, is plotted in a double logarithmic scale as a function of $\Sres$ for random $3$-SAT problems with $N=10^6$ (upper dataset) and random $4$-SAT problems with $N=5\times 10^4$ (lower dataset). The whitening time measured with different $c$ values seems to diverge at the algorithmic threshold, where the residual complexity $\Sres$ vanishes. The lines are power law fits with exponent $\nu=0.281(6)$ for $K\!=\!3$ and $\nu=0.269(5)$ for $K\!=\!4$. Error bars are sem.}
\label{fig6}
\end{figure}

\textbf{Critical exponent for the whitening time divergence} In order to quantify the increase of the whitening time approaching the algorithmic threshold, and inspired by critical phenomena, we check for a power law divergence as a function of $(\alpha_{\rm a}^\text{\tiny BSP}-\alpha)$ or $\Sres$, which are linearly related.
In Fig.\ref{fig6} we plot in a double logarithmic scale the mean whitening time $\tau(c)$ as a function of the residual complexity $\Sres$, for different choices of the fraction $c$ of non-$\star$ variables defining the whitening time.
Data points are fitted via the power law $\tau(c) = A(c) + B(c) \Sres^{-\nu}$, where the critical exponent $\nu$ is the same for all the $c$ values.
Joint interpolations return the following best estimates for the critical exponent: $\nu\!=\!0.281(6)$ for $K\!=\!3$ and $\nu\!=\!0.269(5)$ for $K\!=\!4$, where the uncertainties are only fitting errors.
The two estimate turn out to be compatible within errors, thus suggesting a sort of universality for the critical behavior close to the algorithmic threshold $\alpha_{\rm a}^\text{\tiny BSP}$.

Nonetheless a word of caution is needed since the solutions we are using as starting points for the whitening procedure are atypical solutions (otherwise they would likely contain frozen variables and would not flow to the all-$\star$ configuration under the whitening procedure). 
So, while finding universal critical properties in a dynamical process is definitely a good news, how to relate it to the behavior of the same process on typical solutions it is not obvious (and indeed for the whitening process starting from typical solutions one would expect the naive mean field exponent $\nu=1/2$, which is much larger than the one we are finding).

\section*{Discussion}

We have studied the Backtracking Survey Propagation (BSP) algorithm for finding solutions in very large random $K$-SAT problems and provided numerical evidence that it works much better than any previously available algorithm. That is, BSP has the largest algorithmic threshold known at present.
The main reason for its superiority is the fact that variables can be re-assigned at any time during the run, even at the very end. In other solving algorithms that may look similar, as e.g.\ survey propagation reinforcement \cite{Chavas2005}, re-assignment of variables actually takes place mostly at the beginning of the run, and this is far less efficient in hard problems.
Even doing a lot of helpful backtracking, the BSP running time is still $O(N \log N)$ in the worst case, and thanks to this it can be used on very large problems with millions of constraints.

For $K\!=\!3$ the BSP algorithm finds solutions practically up to the SAT-UNSAT threshold $\alpha_{\rm s}$, while for $K\!=\!4$ a tiny gap to the SAT-UNSAT threshold still remains, but the algorithmic threshold $\alpha_{\rm a}^\text{\tiny BSP}$ seems to be located beyond the rigidity threshold $\alpha_{\rm r}$ in the large $N$ limit.
Beating the rigidity threshold, i.e.\ finding solutions in a region where the majority of solutions belongs to clusters with frozen variables, is hard, but not impossible (while going beyond $\alpha_{\rm f}$ should be impossible). Indeed, even under the assumption that finding frozen solutions takes an exponential time in $N$, very smart polynomial time algorithms can look for a solution in the sub-dominant unfrozen clusters \cite{DallAstaRamezZecchina,LockedCSP}. BSP belongs to this category, as we have shown that all solutions found by BSP have no frozen variables. 

One of the main questions we tried to answer with our extensive numerical simulations is whether BSP is reaching (or approaching closely) the ultimate threshold $\alpha_{\rm a}$ for polynomial time algorithms solving large random $K$-SAT problems.
Under the assumption that frozen solutions cannot be found in polynomial time, such an algorithmic threshold $\alpha_{\rm a}$ would coincide with the freezing transition at $\alpha_{\rm f}$ (i.e.\ when the last unfrozen solution disappears).
Unfortunately for random $K$-SAT the location of $\alpha_{\rm f}$ is not known with enough precision to allow us to reach a definite answer to this question.
It would be very interesting to run BSP on random hypergraph bicoloring problems, where the threshold values are known \cite{bicoloring,COZ2012} and a very recent work has shown that the large deviation function for the number of unfrozen clusters can be computed \cite{SigmaUnfrozen}.

It is worth noticing that the BSP algorithm is easy to parallelize, since most of the operations are local and do not require any strong centralized control. Obviously the effectiveness of a parallel version of the algorithm would largely depend on the topology of the factor graph representing the specific problem: if the factor graph is an expander, then splitting the problem on several cores may require too much inter-core bandwidth, but in problems having a natural hierarchical structure the parallelization may lead to further performance improvements.

The backtracking introduced in the BSP algorithm helps a lot in correcting errors made during the partial assignment of variables and this allows the BSP algorithm to reach solutions at large $\alpha$ values. Clearly we pay the price that a too frequent backtracking makes the algorithm slower, but it seems worth paying such a price to approach the SAT-UNSAT threshold closer than any other algorithm.

A natural direction to improve this class of algorithms would be to used {\bf biased} marginals focusing on solutions which are easier to be reached by the algorithm itself. For example in the region $\alpha>\alpha_{\rm r}$ the measure is concentrated on solutions with frozen variables, but these can not be really reached by the algorithm. The backtracking thus intervenes and corrects the partial assignment until a solution with unfrozen variables is found by chance. If the marginals could be computed from a new biased measure which is concentrated on the unfrozen clusters, this could make the algorithm go immediately in the right direction and much less backtracking would be hopefully needed.

\section*{Methods}

\textbf{Survey Inspired Decimation (SID)}
A detailed description of the SID algorithm can be found in Refs.~\cite{MPZ2002,MezardZecchina2003,BMZ2005}.
The SID algorithm is based on the survey propagation (SP) equations derived by the cavity method \cite{MPZ2002,MezardZecchina2003}, that can be written in a compact way as
\begin{eqnarray}
\hat{m}_{a\to i} &=& \prod_{j\in\partial_a\setminus i} m_{j \to a}\,,\\
m_{i\to a} &=& \frac{\pi^-_{ia}(1-\pi^+_{ia})}{1-\pi^+_{ia}\pi^-_{ia}}\,,\\
\text{with}\quad\pi^\pm_{ia} &=& 1-\prod_{b \in \partial^\pm_{ia}} (1-\hat{m}_{b\to i})\,,
\end{eqnarray}
where $\partial_a$ is the set of variables in clause $a$, and $\partial^+_{ia}$ (resp.\ $\partial^-_{ia}$) is the set of clauses containing $x_i$, excluding $a$ itself, satisfied (resp.\ not satisfied) when the variable $x_i$ is assigned to satisfy clause $a$.

The interpretation of the SP equations is as follows: $\hat{m}_{a\to i}$ represents the fraction of clusters where clause $a$ is satisfied solely by variable $x_i$ (that is, $x_i$ is frozen by clause $a$), while $m_{i\to a}$ is the fraction of clusters where $x_i$ is frozen to an assignment not satisfying clause $a$.

The SP equations impose $2KM$ self-consistency conditions on the $2KM$ variables $\{m_{i\to a},\hat{m}_{a\to i}\}$ living on the edges of the factor graph \cite{factorGraph}, that are solved in an iterative way, leading  to a message passing algorithm (MPA) \cite{bookMezardMontanari}, where outgoing messages from a factor graph node (variable or clause) are functions of the incoming messages. Once the MPA reaches a fixed point $\{m^\star_{i\to a},\hat{m}^\star_{a\to i}\}$ that solves the SP equations, the number of clusters can be estimated via the complexity
\begin{align}
&\Sigma = \log\mathcal{N}_\text{clus} = \sum_i \Sigma_i + \sum_a (1-K_a) \Sigma_a\,,\\
&\Sigma_a = \log\Big(1\!-\!\prod_{j\in\partial_a} m^\star_{j\to a}\Big)\,,\quad
\Sigma_i = \log(1\!-\pi^+_i\pi^-_i)\\
&\text{with}\quad \pi^\pm_i=1-\prod_{b \in \partial^\pm_i} (1-\hat{m}^\star_{b\to i})\,,
\end{align}
where $K_a$ is the length of clause $a$ (initially $K_a=K$) and $\partial^+_i$ (resp.\ $\partial^-_i$) is the set of clauses satisfied by setting $x_i=1$ (resp.\ $x_i=-1$). The SP fixed point messages also provide information about the fraction of clusters where variable $x_i$ is forced to be positive ($w_i^+$), negative ($w_i^-$) or not forced at all ($1-w_i^+-w_i^-$)
\begin{equation}
w_i^\pm = \frac{\pi_i^\pm(1-\pi_i^\mp)}{1-\pi_i^+\pi_i^-}\,.
\end{equation}

The SID algorithm then proceed by assigning variables (decimation step). According to SP equations, assigning a variable $x_i$ to its most probable value (i.e., setting $x_i=1$ if $w_i^+>w_i^-$ and viceversa), the number of clusters gets multiplied by a factor, called {\bf bias}
\begin{equation}
b_i = 1-\min(w_i^+,w_i^-)\,.
\end{equation}
With the aim of decreasing the lesser the number of cluster and thus keeping the largest the number of solutions in each decimation step, SID assigns/decimate variables with the largest $b_i$ values. In order to keep the algorithm efficient, at each step of decimation a small fraction $f$ of variables is assigned, such that in $O(\log N)$ steps of decimation a solution can be found.

After each step of decimation, the SP equations are solved again on the subproblem, which is obtained by removing satisfied clauses and by reducing clauses containing a false literal (unless a zero-length clause is generated, and in that case the algorithm returns a failure). The complexity and the biases are updated according to the new fixed point messages, and a new decimation step is performed.

The main idea of the SID algorithm is that fixing variables which are almost certain to their most probable value, one can reduce the size of the problem without reducing too much the number of solutions. The evolution of the complexity $\Sigma$ during the SID algorithm can be very informative \cite{ParisiRemarks2003}.
Indeed it is found that, if $\Sigma$ becomes too small or negative, the SID algorithm is likely to fail, either because the iterative method for solving the SP equations no longer converges to a fixed point or because a contradiction is generated by assigning variables. In these cases the SID algorithm returns a failure.
On the contrary, if $\Sigma$ always remains well positive, the SID algorithm reduces so much the problem, that eventually a trivial SP fixed point, $m^\star_{i\to a}=\hat{m}^\star_{a\to i}=0$, is reached. This is a strong hint that the remaining subproblem is easy and the SID algorithm tries to solve it by WalkSat \cite{WalkSat}. 

A careful analysis of the SID algorithm for random $3$-SAT problems of size $N=O(10^5)$ shows that the algorithmic threshold achievable by SID is $\alpha_{\rm a}^\text{\tiny SID}=4.2525$ \cite{ParisiRemarks2003}, which is close, but definitely smaller than the SAT-UNSAT threshold $\alpha_{\rm s}=4.2667$.

The running time of the SID algorithm experimentally measured is $O(N \log(N))$ \cite{BMZ2005}.

\textbf{Backtracking Survey Propagation (BSP)}
Willing to improve the SID algorithm to find solutions also in the region $\alpha_{\rm a}^\text{\tiny SID}<\alpha<\alpha_{\rm s}$, one has to change the way variables are assigned. The fact the SID algorithm assigns each variable only once is clearly a strong limitation, especially in a situation where correlations between variables becomes extremely strong and long-ranged. In difficult problems it can easily happen that one realizes that a variable is taking the wrong value only after having assigned some of its neighbours variables. However the SID algorithm is not able to solve this kind of frustrating situations.

The BSP algorithms \cite{ParisiBSP2003} tries to solve this kind of problematic situations by introducing a new {\bf backtracking} step, where a variable already assigned can be released and eventually re-assigned in a future decimation step.
It is not difficult to understand when it is worth releasing a variable.
The bias $b_i$ in terms of the SP fixed point messages $\{\hat{m}^\star_{a\to i}\}_{a\in\partial_i}$ arriving in $i$ can be computed also for a variable $x_i$ already assigned: if the bias $b_i$, that was large at the time the variable $x_i$ was assigned, gets strongly reduces by the effect of assigning other variables, then it is likely that releasing the variable $x_i$ may be beneficial in the search for a solution.
So both the variables to be fixed in the decimation step and the variables to be released in the backtracking step are chosen according to their biases $b_i$: the variables to be fixed have the largest biases and the variables to be released have the smallest biases.

The BSP algorithm then proceeds similarly to SID, by alternating the iterative solution to the SP equations and a step of decimation or backtracking on a fraction $f$ of variables in order to keep the algorithm efficient (in all our numerical experiments we have used $f=10^{-3}$).
The choice between a decimation or a backtracking step is taken according to a stochastic rule (unless there are no variables to unset), where the parameter $r\in[0,1)$ represents the ratio between backtracking steps to decimation steps.
Obviously for $r=0$ we recover the SID algorithm, since no backtracking step is ever done. Increasing $r$ the algorithm becomes slower by a factor $1/(1-r)$, because variables are reassigned on average $1/(1-r)$ times each before the BSP algorithm reaches the end, but its complexity remains at most $O(N \log N)$ in the problem size.

The BSP algorithm can stop for the same reasons the SID algorithm does: either the SP equations can not be solved iteratively or the generated subproblem has a contradiction. Both cases happen when the complexity $\Sigma$ becomes too small or negative.
On the contrary if the complexity remain always positive the BSP eventually generate a subproblem where all SP messages are null and on this subproblem WalkSat is called.

\subsection*{Data Availability Statement}
The numerical codes used in this study and the data that support the findings are available from
the corresponding author upon request.

\section*{Acknowledgements}
We thank K.~Freese, R.~Eichhorn and E.~Aurell for useful discussions. This research has been supported by the Swedish Science Council through grant 621-2012-2982 and by the European Research Council
(ERC) under the European UnionÕs Horizon 2020 research and innovation
programme (grant agreement No [694925]).

\section*{Author contributions}
All authors contributed to all aspects of this work.

\section*{Additional information}
\textbf{Competing financial interests:} The authors declare no competing financial interests.

\end{document}